\documentclass[aps,prd,onecolumn,nofootinbib,superscriptaddress, 10.5pt]{revtex4-2}
\usepackage{bm}
\usepackage{multirow}
\usepackage{tikz}
\usepackage{amsmath, amssymb, amstext}
\usepackage{graphicx}
\usepackage{hyperref}
\usepackage{physics}
\usepackage{enumitem}
\usepackage{geometry}
\usepackage{float}
\usepackage{xcolor}
\usepackage{soul}
\begin{document}

\title{Primordial Black Hole Formation in Rastall Gravity: Shifted Collapse Threshold and Exponential Abundance Sensitivity}

\author{Mayukh R. Gangopadhyay \\
\small Department of Physics, School of Advanced Sciences, \\
\small Vellore Institute of Technology (VIT), Chennai Campus, \\
\small Chennai 600127, Tamil Nadu, India \\
\small \texttt{mayukhraj.g@vit.ac.in}}

\maketitle

\begin{center}\textbf{Abstract}\end{center}
Primordial black holes formed in the early universe are compelling candidates for dark matter. We investigate their production in Rastall gravity, a modification of general relativity that introduces a non-minimal coupling between matter and geometry through the non-conservation of the energy-momentum tensor. Analyzing cosmological perturbations during radiation domination, we demonstrate that the Rastall parameter fundamentally alters the collapse dynamics, modifying the growth of density fluctuations, the critical threshold for black hole formation, and the fluctuation amplitude at horizon crossing if we consider the nearly conformal plasma. Current cosmological constraints from Big Bang Nucleosynthesis, the cosmic microwave background, and large-scale structure restrict the Rastall parameter to small values, yet within this allowed range PBH production can be altered by orders of magnitude compared to general relativity. Our results suggest that PBH formation provides a sensitive and independent probe of perturbation-level modifications in Rastall gravity, complementary to large-scale structure and CMB tests.


\section{Introduction}

Primordial black holes (PBHs) remain among the most intriguing objects in modern cosmology. Formed by the gravitational collapse of large-amplitude density perturbations re-entering the Hubble horizon during radiation domination, they have attracted sustained attention as dark matter candidates and as probes of early-universe physics that are otherwise inaccessible \cite{Carr2005, Carr2020, Green2021, Khlopov2010, Carr:2020gox, Carr:2009jm}. The connection to gravitational wave astronomy has added a further observational handle \cite{Bagui2025, Bird2016, Sasaki2016, Clesse2017, Nakamura1997, Ioka1998}. In the standard GR picture, however, producing any appreciable PBH abundance requires a large enhancement of the primordial curvature power spectrum at small scales---an enhancement that typically calls for non-trivial, often fine-tuned, inflationary model building \cite{Motohashi2017, Ballesteros2018, GarciaBellido1996, Kawasaki1998, Yokoyama1998, Gangopadhyay2022, Nandi2025, Correa2022, FineTuning2024a, FineTuning2024b}. PBH populations, once formed, can be constrained through microlensing, CMB spectral distortions, and other observational channels \cite{Stojkovic:2004hz, Stojkovic:2005zh, Dai:2024guo, Dong:2015yjs}.

Modified gravity theories have long been discussed as a way to alter PBH production rates, since any change to the collapse dynamics or to the primordial power spectrum carries through exponentially into the final abundance \cite{Capozziello2011, Nojiri2017, Clifton2012, Sotiriou2010, DeFelice2010, PBHModGrav2021, BarrowMajumdar2003}. In this paper we focus on Rastall gravity \cite{Rastall1972, Rastall1976}. PBH formation in various modified gravity frameworks has recently attracted attention \cite{RastallPBH2025}; the present work provides a complementary perturbation-level analysis focused on Rastall gravity. The theory is built on a single modification: instead of demanding covariant conservation of the energy-momentum tensor, Rastall postulated a direct coupling to the gradient of the Ricci scalar,

\begin{equation}
\nabla_\mu T^{\mu\nu} = \lambda \nabla^\nu R, \label{eq:rastall}
\end{equation}

where $\lambda$ is the Rastall parameter and $R$ is the Ricci scalar, while preserving the Bianchi identity $\nabla_\mu G^{\mu\nu}=0$. This leads to the
modified field equations
\begin{equation}
G_{\mu\nu}
=
\kappa
\left(
T_{\mu\nu}
-
\frac{\lambda}{1-4\lambda} g_{\mu\nu} T
\right),
\end{equation}
which reduce to general relativity (GR) in the limit $\lambda \to 0$ \cite{daRocha2023, Fabris2022, Darabi2018, Moradpour2017, Visser2018, Bronnikov2017, Smalley1984, Lindblom1982}.

There is a long-standing debate in the literature over whether Rastall gravity is genuinely distinct from GR or merely a redefinition of the matter content \cite{Visser2018}. The field equations can indeed be recast as standard Einstein equations with an effective stress tensor,
\begin{equation}
G_{\mu\nu} = \kappa \, T^{\rm (eff)}_{\mu\nu},
\end{equation}
where
\begin{equation}
T^{\rm (eff)}_{\mu\nu}
=
T_{\mu\nu}
-
\frac{\lambda}{1-4\lambda} g_{\mu\nu} T.
\end{equation}
From this point of view, the theory looks like GR with a non-standardly conserved matter sector \cite{Visser2018}.

The resolution depends on what one takes $T_{\mu\nu}$ to represent. If it is the physical matter tensor of a radiation fluid, then the modified conservation law implies a genuine exchange of energy-momentum between matter and geometry that is absent in GR---and the perturbation dynamics are accordingly altered. That is the interpretation we adopt here.

The background evolution of a pure radiation fluid turns out to be unaffected: for $w=1/3$ one finds $T=0$, so the trace coupling vanishes at zeroth order and the Friedmann equations reduce to their GR form. Perturbations, however, are sensitive to the trace structure at first order, and the modified conservation equation leaves a non-trivial imprint. This is the key feature that makes PBH formation during radiation domination a useful probe---any difference from GR must come entirely from the perturbation sector, not from a modified expansion history.

No additional propagating degrees of freedom are introduced at the linearised level: there are no extra tensor polarisations around Minkowski space. The distinguishable physics enters purely through the modified matter--geometry coupling, which is why early-universe probes that are sensitive to perturbation dynamics---like PBH formation---are particularly well-suited to testing the theory.

The Rastall framework has been studied fairly widely. Black hole solutions have been examined by a number of groups \cite{Heydarzade2018, Heydarzade2017, Li2022, Xu2022, Kumar2019, Ma2017, Zangeneh2020, Lobo2020}, with modifications to horizon thermodynamics reported. On the cosmological side there is work on background evolution \cite{Batista2012, Batista2013, Fabris2012, Das2018, Oliveira2015, AbdelRahman1997, Lin2020, Khadekar2019}, perturbation dynamics \cite{Salti2020, Salti2021, Fabris2000, Fabris2012b, Moradpour2018, Licata2017, Corda2018, Hansraj2019}, structure formation, and gravitational wave propagation \cite{Salti2020, Salti2021, Moradpour2018, Licata2017, Corda2018}. The specific question of PBH formation in Rastall gravity has received less attention, and a systematic perturbation-level treatment appears not to have been carried out.

\section{Rastall Gravity: Fundamentals and Background Evolution}

Combining the Bianchi identity $\nabla^\mu G_{\mu\nu} = 0$ with the non-conservation postulate \eqref{eq:rastall} gives the Rastall field equations:

\begin{equation}
G_{\mu\nu} = R_{\mu\nu} - \frac{1}{2}g_{\mu\nu}R = \kappa \left(T_{\mu\nu} - \frac{\lambda}{1-4\lambda}g_{\mu\nu}T\right), \label{eq:field}
\end{equation}

where $\kappa = 8\pi G$ and $T = T^\mu_\mu$. The constraint $\lambda \neq 1/4$ is required to avoid a singularity; GR is recovered when $\lambda \to 0$.

For a perfect fluid with $p = w\rho$:

\begin{equation}
T_{\mu\nu} = (\rho + p)u_\mu u_\nu + p g_{\mu\nu}, \quad T = -\rho + 3p = (3w-1)\rho. \label{eq:stress}
\end{equation}

Taking a flat FLRW metric:

\begin{equation}
ds^2 = -dt^2 + a^2(t)\delta_{ij}dx^i dx^j, \label{eq:flrw}
\end{equation}

where $H = \dot{a}/a$. Substituting into \eqref{eq:field} gives:

\begin{align}
3H^2 &= \kappa \left[1 - \frac{\lambda(1-3w)}{1-4\lambda}\right]\rho, \label{eq:f1} \\
2\dot{H} + 3H^2 &= -\kappa \left[w - \frac{\lambda(1-3w)}{1-4\lambda}\right]\rho. \label{eq:f2}
\end{align}

The time component of \eqref{eq:rastall} on the FLRW background gives:

\begin{equation}
\dot{\rho} + 3H(\rho + p) = \lambda \dot{R}, \quad R = 6(2H^2 + \dot{H}). \label{eq:cons}
\end{equation}

For a radiation fluid, $w = 1/3$ and $1-3w = 0$, so Eqs.~\eqref{eq:f1}--\eqref{eq:f2} reduce to:

\begin{align}
3H^2 &= \kappa\rho, \label{eq:r1} \\
2\dot{H} + 3H^2 &= -\frac{\kappa\rho}{3}. \label{eq:r2}
\end{align}

The background equations are therefore indistinguishable from GR:

\begin{align}
H^2 &= \frac{8\pi G}{3}\rho, \label{eq:r3} \\
\dot{H} &= -2H^2, \label{eq:r4} \\
\rho &\propto a^{-4}, \quad a \propto t^{1/2} \propto \tau, \label{eq:r5}
\end{align}

where $\tau$ is conformal time, $d\tau = dt/a$. For radiation ($p = \rho/3$) the conservation equation \eqref{eq:cons} reads:

\begin{equation}
\dot{\rho} + 4H\rho = \lambda \dot{R}. \label{eq:consrad}
\end{equation}

The Ricci scalar follows from the Friedmann equations:

\begin{equation}
R = 6(2H^2 + \dot{H}) = 6(2H^2 - 2H^2) = 0. \label{eq:ricci}
\end{equation}

Since $R = 0$ and $\dot{R} = 0$, Eq.~\eqref{eq:consrad} reduces to the standard result $\dot{\rho} + 4H\rho = 0$. The background is therefore identical to GR throughout radiation domination.

Differences with GR emerge only at the perturbation level, and only when the plasma deviates slightly from exact conformality---as discussed in Section~\ref{activation} below. This is actually what makes PBH formation a clean test of the theory: since the background is unchanged, any Rastall signature must originate in the perturbation dynamics.

\subsection{Activation of Rastall Corrections in a Nearly Conformal Plasma}
\label{activation}
For an exactly conformal fluid with $p=\rho/3$, the trace vanishes identically:
\begin{equation}
T \equiv T^\mu_{\ \mu} = -\rho + 3p = 0.
\end{equation}
In that case the modified conservation law
\begin{equation}
\nabla_\mu T^{\mu\nu} = \lambda \nabla^\nu R
\end{equation}
reduces to the standard one, and Rastall gravity is dynamically indistinguishable from GR at every order in perturbation theory.

The primordial plasma is not, of course, exactly conformal. Near the QCD epoch---precisely the temperature range relevant for horizon re-entry of the scales of interest for PBH formation---the trace anomaly and deviations from ideal fluid behaviour generate a small but non-zero $T$. We parametrise this as
\begin{equation}
p = \left(\frac{1}{3} - \varepsilon \right)\rho,
\qquad \varepsilon \ll 1,
\end{equation}
which implies
\begin{equation}
T = -\rho + 3p = -3\varepsilon \rho.
\end{equation}

We adopt the effective equation of state parametrised by Saikawa and Shirai \cite{Saikawa2018} (see also Husdal \cite{Husdal2016}), which provides a smooth fit to lattice QCD results for $w(T)$ and the effective degrees of freedom $g_*(T)$ across the QCD crossover. The conformality-breaking parameter $\varepsilon(T) \equiv \tfrac{1}{3} - w(T)$ is tabulated in Table~\ref{tab:eos} for the temperature range relevant to stellar-mass PBH formation.

\begin{table}[h!]
\centering
\begin{tabular}{ccccc}
\hline
$T$ [MeV] & $w(T)$ & $\varepsilon(T) = \tfrac{1}{3}-w$ & $g_*(T)$ & $M_H$ [$M_\odot$] \\
\hline
80  & 0.331 & 0.002 & 10.75 & 40.6 \\
100 & 0.314 & 0.019 & 10.75 & 26.0 \\
120 & 0.253 & 0.081 & 11.09 & 17.7 \\
130 & 0.202 & 0.132 & 13.68 & 13.6 \\
140 & 0.148 & 0.185 & 15.43 & 11.1 \\
150 & 0.110 & 0.224 & 17.25 & 9.1  \\
155 & 0.100 & 0.233 & 17.25 & 7.8  \\
160 & 0.099 & 0.234 & 24.88 & 6.7 \\
170 & 0.120 & 0.214 & 34.66 & 5.0  \\
180 & 0.162 & 0.171 & 40.80 & 4.1  \\
200 & 0.254 & 0.079 & 51.25 & 3.0  \\
300 & 0.333 & 0.000 & 61.75 & 1.2  \\
\hline
\end{tabular}
\caption{Equation of state parameter $w(T)$, conformality-breaking parameter $\varepsilon(T) = \tfrac{1}{3}-w(T)$, effective degrees of freedom $g_*(T)$, and horizon mass $M_H(T)$ across the QCD crossover, computed from the Saikawa--Shirai parametrisation \cite{Saikawa2018}. The bold entry marks the peak of $\varepsilon$ near $T\approx 155$--$160$~MeV. The horizon mass column shows the PBH mass scale probed at each temperature.}
\label{tab:eos}
\end{table}

PBHs forming at the QCD epoch span the mass range $M \sim 1$--$30\,M_\odot$ (corresponding to $T \sim 150$--$500$~MeV as read from Table~\ref{tab:eos}), which coincides precisely with the temperature window where $\varepsilon$ is largest ($\varepsilon_{\rm max} \simeq 0.23$ at $T\simeq 158$~MeV). This is the mass range relevant to LIGO/Virgo merger events and to a fraction of PBH dark matter; it is also the window where the Rastall corrections computed in this paper are most significant.

The leading Rastall correction to the perturbation equation scales as $\lambda\varepsilon$ (see the $\delta T \propto \varepsilon\,\delta\rho$ term above). Varying $\varepsilon$ across its range $0$--$0.23$ therefore changes the effective size of the Rastall modification by the same factor. At $T \gg 200$~MeV or $T \ll 100$~MeV, where $\varepsilon \to 0$, the Rastall corrections are strongly suppressed and the theory is observationally indistinguishable from GR at perturbation level. The effect is maximal near the QCD crossover. In deriving the master equation (Section~4) we absorb the $\varepsilon$ factor into the $\mathcal{O}(1)$ coefficient $\alpha_H$; the representative value $\alpha_\sigma \approx -4.1$ quoted in Section~4.2.2 already incorporates this suppression.

The background remains radiation-dominated to good approximation, but the trace no longer vanishes. At linear order,
\begin{equation}
\delta T = -\delta\rho + 3\delta p = -3\varepsilon\,\delta\rho,
\end{equation}
so the modified conservation law contributes at $\mathcal{O}(\varepsilon)$ in the scalar perturbation equation, with corrections scaling as $\lambda\varepsilon$. In deriving the master equation (Section~4) we absorb this $\varepsilon$ factor into the $\mathcal{O}(1)$ coefficient $\alpha_H$, treating the full Rastall correction as $\lambda\times\mathcal{O}(1)$ throughout.

\section{Scalar Perturbations and Density Contrast Evolution}

We work in the Newtonian gauge ($\Phi = \Psi$, no anisotropic stress):

\begin{equation}
ds^2 = -(1 + 2\Phi)dt^2 + a^2(t)(1 - 2\Phi)\delta_{ij}dx^i dx^j. \label{eq:metricpert}
\end{equation}

The perturbed stress tensor components are:

\begin{align}
\delta T^0_0 &= -\delta\rho, \label{eq:pertT00} \\
\delta T^0_i &= (\rho + p)\partial_i v, \label{eq:pertT0i} \\
\delta T^i_j &= \delta p \, \delta^i_j, \label{eq:pertTij}
\end{align}

where $\delta\rho$ is the density perturbation, $\delta p = c_s^2\delta\rho$ is the pressure perturbation (with $c_s^2 = w$ for adiabatic perturbations), and $v$ is the velocity potential.

Varying \eqref{eq:field}, the 00-component in Fourier space gives:

\begin{equation}
-\frac{2k^2}{a^2}\Phi + 6H(\dot{\Phi} + H\Phi) = \kappa \left[\left(1 - \frac{4\lambda}{1-4\lambda}c_s^2\right)\delta\rho\right]. \label{eq:pert00}
\end{equation}

The 0i-component:

\begin{equation}
2\partial_i(\dot{\Phi} + H\Phi) = \kappa(\rho + p)\partial_i v. \label{eq:pert0i}
\end{equation}

The perturbed conservation law $\delta(\nabla_\mu T^{\mu\nu}) = \lambda \nabla^\nu \delta R$ gives, for $\nu=0$:

\begin{equation}
\delta(\nabla_\mu T^{\mu 0}) = -\lambda \partial_0 \delta R. \label{eq:pertcons0}
\end{equation}

The left-hand side expands as:

\begin{equation}
\delta(\nabla_\mu T^{\mu 0}) = \dot{\delta\rho} + 3H(\delta\rho + \delta p) + (\rho + p)\left(3\dot{\Phi} - \frac{k^2}{a^2}v\right). \label{eq:pertenerg}
\end{equation}

For $\nu = i$:

\begin{equation}
\delta(\nabla_\mu T^{\mu i}) = \frac{\lambda}{a^2}\partial^i \delta R. \label{eq:pertconsi}
\end{equation}

giving:

\begin{equation}
\delta(\nabla_\mu T^{\mu i}) = (\rho + p)(\dot{v} + 4Hv) + \partial^i \delta p + (\rho + p)\partial^i \Phi. \label{eq:pertmom}
\end{equation}

The first-order perturbed Ricci scalar in Fourier space is:

\begin{equation}
\delta R = \frac{2k^2}{a^2}\Phi + 6\ddot{\Phi} + 12H\dot{\Phi} + 6(2H^2 + \dot{H})\Phi. \label{eq:pertR}
\end{equation}

\section{The Master Equation for the Density Contrast}

We now combine the above to obtain a single second-order equation for $\delta \equiv \delta\rho/\rho$. Starting from the 0i-component \eqref{eq:pert0i}:

\begin{equation}
v = \frac{2(\dot{\Phi} + H\Phi)}{\kappa(\rho + p)}. \label{eq:vphi}
\end{equation}
Now,
from the 00-component \eqref{eq:pert00}, for radiation with $c_s^2 = 1/3$ and working to linear order in $\lambda$ (the observationally allowed regime, $|\lambda| \ll 1$), we obtain:

\begin{equation}
\delta\rho \approx -\frac{1}{\kappa}\left(1 + \frac{4\lambda}{3}\right)\left(\frac{2k^2}{a^2}\Phi + 6H(\dot{\Phi} + H\Phi)\right). \label{eq:deltarho}
\end{equation}
Using $\kappa\rho = 3H^2$ from \eqref{eq:r1}, this can be inverted to give $\Phi$ in terms of $\delta$. To linear order in both perturbations and $\lambda$:

\begin{equation}
\Phi \approx -\frac{a^2}{2k^2}\left[\frac{\kappa\rho}{3H^2}\left(1 - \frac{4\lambda}{3}\right)\delta + 6H(\dot{\Phi} + H\Phi)\right]. \label{eq:phidelta}
\end{equation}

Substituting \eqref{eq:vphi}, \eqref{eq:deltarho} and the perturbed conservation equations \eqref{eq:pertmom}, \eqref{eq:pertconsi}:

\begin{equation}
(\rho + p)(\dot{v} + 4Hv) + \partial^i\delta p + (\rho + p)\partial^i\Phi = \frac{\lambda}{a^2}\partial^i\delta R. \label{eq:momfinal}
\end{equation}

Using the standard radiation-domination relations ($\rho + p = 4\rho/3$, $\dot{H} = -2H^2$, $c_s^2 = 1/3$) and switching to conformal time ($a \propto \tau$, $\mathcal{H} = 1/\tau$), keeping only terms linear in $\lambda$:

\begin{equation}
\Phi'' + \frac{4}{\tau}\Phi' + \frac{k^2}{3}\Phi = \lambda\left(\frac{6}{\tau^2}\Phi'' + \frac{24}{\tau^3}\Phi' - \frac{4k^2}{3\tau^2}\Phi\right) + \mathcal{O}(\lambda^2). \label{eq:phievol}
\end{equation}

Using the relation between $\delta$ and $\Phi$ derived from \eqref{eq:deltarho} in conformal time:

\begin{equation}
\delta \approx -\frac{2}{3}\left(1 + \frac{4\lambda}{3}\right)\left(\frac{k^2\tau^2}{a^2}\Phi + 3\tau\Phi' + 3\Phi\right). \label{eq:deltaphi}
\end{equation}

Differentiating \eqref{eq:deltaphi} twice and substituting into \eqref{eq:phievol}, after some algebra one finds:

\begin{equation}
\boxed{\ddot{\delta} + 2H\dot{\delta} - \left[\frac{3}{2}H^2 - \frac{k^2}{3a^2} + \lambda\left(6H^2 - \frac{4k^2}{3a^2}\right)\right]\delta = 0.} \label{eq:master}
\end{equation}

This is the central equation of the perturbation analysis. Compared to the GR equation, both the effective gravitational driving term and the effective sound speed carry $\lambda$-dependent corrections.

\subsection{Solution During Radiation Domination}

In conformal time during radiation domination it is convenient to work with $x \equiv k\tau = k/(aH)$. The master equation becomes:

\begin{equation}
\frac{d^2\delta}{dx^2} + \frac{2}{x}\frac{d\delta}{dx} + \left[\frac{1}{3} - \frac{3}{2x^2} - \frac{4\lambda}{3}\left(1 - \frac{3}{x^2}\right)\right]\delta = 0. \label{eq:masterx}
\end{equation}

\subsubsection{Super-horizon solution ($x \ll 1$)}

For $x \ll 1$ we write $\delta(x) = \delta_0[1 + \alpha x^2 + \mathcal{O}(x^4)]$ and match coefficients:

\begin{equation}
\delta(x) \approx \delta_0\left[1 + \left(\frac{3}{8} - \frac{2\lambda}{3}\right)x^2 + \mathcal{O}(x^4)\right]. \label{eq:superhor}
\end{equation}

The Rastall correction shifts the super-horizon growth rate but remains $\mathcal{O}(x^2)$ and is small before horizon crossing.

\subsubsection{Horizon crossing ($x = 1$)}

Matching to the numerical solution at $x=1$:

\begin{equation}
\delta_H(k) = \frac{4}{9}(1 + \alpha_H \lambda)\mathcal{P}_\mathcal{R}^{1/2}(k), \label{eq:horcross}
\end{equation}

where $\alpha_H$ is a dimensionless coefficient set by the perturbation matching and $\mathcal{P}_\mathcal{R}(k)$ is the primordial curvature power spectrum. Note that $\alpha_\sigma \equiv \alpha_H$ in Section~5.2: it is a purely linear-theory quantity, entirely independent of the nonlinear coefficient $\gamma$.

The coefficient $\alpha_\sigma$ is extracted directly from the numerical solution of Eq.~(\ref{eq:masterx}) by reading off the horizon-crossing amplitude ratio
\begin{equation}
\frac{\delta_H(\lambda)}{\delta_H^{\rm GR}} = 1 + \alpha_\sigma\,\lambda + \mathcal{O}(\lambda^2),
\label{eq:alphasigma_def}
\end{equation}
at $x = 1$. From Figure~\ref{fig4} and the underlying numerical solutions one finds $\delta_H/\delta_H^{\rm GR} \in \{1.48, 1.22, 1.00, 0.81, 0.65\}$ for $\lambda \in \{-0.10, -0.05, 0.00, +0.05, +0.10\}$, giving:
\begin{equation}
\alpha_\sigma \approx -4.1 \qquad (\text{representative value}; \; |\alpha_\sigma| = \mathcal{O}(1)),
\label{eq:alphasigma_value}
\end{equation}
with a mild variation of $\pm 0.6$ across the range $|\lambda| \leq 0.10$ due to higher-order corrections. The sign is unambiguously negative: positive $\lambda$ suppresses $\delta_H$ and hence the variance $\sigma(M,\lambda)$. With this value the $\gamma=0$ conservative abundance ratio Eq.~(\ref{eq:enhancement_gamma0}) gives $\beta/\beta_{\rm GR} \approx \exp[+4.1\,(\nu_c^{\rm GR})^2\,|\lambda|]$ for $\lambda < 0$, an enhancement exceeding $10^4$ for $\nu_c^{\rm GR} = 7$ and $\lambda = -0.05$.
The numerical solution of the full master equation~(\ref{eq:masterx}), confirming this analytic result and displaying the sub-horizon phase shift, is presented as Figure~\ref{fig4} in Section~\ref{sec:fig4}.


\section{PBH Formation Criterion in Rastall Gravity}

\subsection{Critical Density Threshold for Collapse}

Whether a given overdensity collapses into a black hole or disperses is determined by the critical threshold $\delta_c$.
In general relativity, numerical simulations \cite{Musco2013, Musco2019, Musco2020, Escriva2020, Escriva2024, Shibata1999} have demonstrated that the collapse threshold depends sensitively on the shape of the perturbation profile. For broad (quasi-Gaussian) profiles the canonical value $\delta_c^{\rm GR} \simeq 0.414$ is obtained, while narrower profiles can shift this value by $\pm 0.025$ or more \cite{Musco2020,Escriva2020}. The value $\delta_c^{\rm GR} \simeq 0.414$ is therefore adopted here as a representative benchmark for broad profiles, and the present analysis is correspondingly limited to this class of perturbation shapes. A profile-independent analysis would require repeating the perturbation matching of Section~4 for each profile, which we leave to future work.

In Rastall gravity the threshold shifts because both the effective gravity and the pressure are modified. We estimate $\delta_c(\lambda)$ using two approaches, working throughout to linear order in $\lambda$ as required by observational bounds.

\subsubsection{Approach 1: Jeans scale modification}
\begin{figure}[htb!]
\centering
\includegraphics[width=0.99\textwidth]{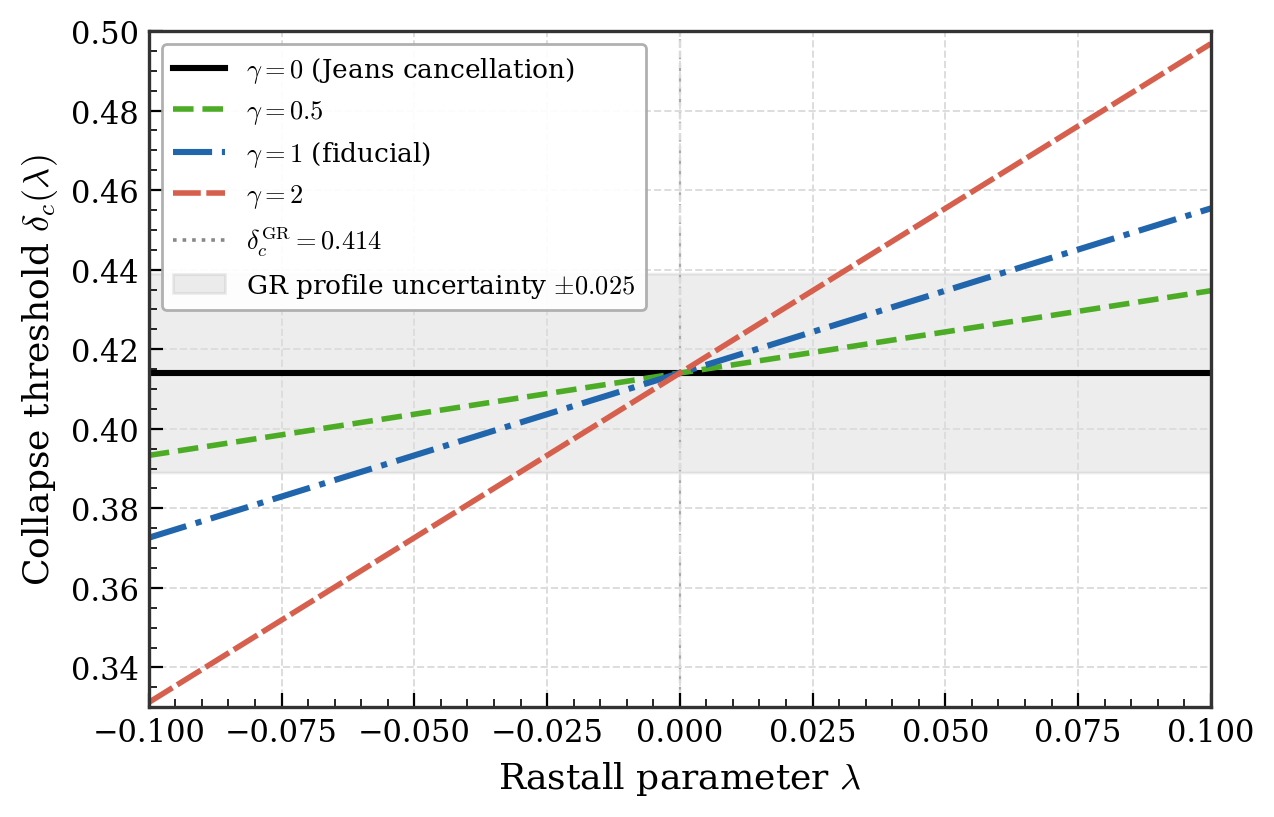}
\caption{Collapse threshold $\delta_c(\lambda)=\delta_c^{\rm GR}(1+\gamma\lambda)$ as a function of the Rastall parameter~$\lambda$, shown for four representative values $\gamma\in\{0,0.5,1,2\}$. The coefficient $\gamma$ is a free phenomenological parameter. The $\gamma=0$ curve (flat, solid black) represents the case in which Rastall corrections to the threshold cancel exactly---as established by the Jeans analysis below---while larger $\gamma$ reflects increasing nonlinear sensitivity. The gray dotted line marks $\delta_c^{\rm GR}=0.414$ and the shaded band shows the $\pm 0.025$ profile-shape uncertainty from GR simulations \cite{Musco2020,Escriva2020,Escriva2024}. For $\gamma>0$: positive $\lambda$ raises the threshold, negative $\lambda$ lowers it.}
\label{fig1}
\end{figure}
Reading off from \eqref{eq:master} by comparison with the standard dispersion form $\ddot{\delta} + 2H\dot{\delta} - (4\pi G_{\rm eff}\rho - c_s^2 k^2/a^2)\delta = 0$, we identify:

\begin{align}
4\pi G_{\text{eff}}\rho &= \frac{3}{2}H^2 - 6\lambda H^2 = \frac{3}{2}H^2(1 - 4\lambda), \label{eq:Geff} \\
c_s^2(\lambda) &= \frac{1}{3} - \frac{4\lambda}{3}. \label{eq:cs2}
\end{align}

Setting $\omega = 0$:

\begin{equation}
k_J^2 = \frac{4\pi G_{\text{eff}}\rho}{c_s^2(\lambda)} a^2 = \frac{\frac{3}{2}H^2(1 - 4\lambda)}{\frac{1}{3}(1 - 4\lambda)} a^2 = \frac{9}{2}H^2 a^2. \label{eq:jeans}
\end{equation}

The $\lambda$ dependence cancels exactly in the ratio. The Jeans scale is unchanged from GR---a non-trivial consequence of the fact that the modifications to $G_{\rm eff}$ and $c_s^2$ carry the same factor $(1-4\lambda)$ and therefore cancel when forming their ratio.

The Jeans wavenumber $k_J$ is a \emph{linear} quantity---it marks onset of gravitational instability in the linearised equation. The PBH threshold $\delta_c$ is \emph{nonlinear}: it encodes the full dynamics of collapse against pressure gradients, which is why even in GR, $\delta_c^{\rm GR} \simeq 0.414$ differs substantially from the linear Jeans value $\delta_{\rm Jeans} \simeq 1/3$ \cite{Musco2013,Musco2020,Escriva2020}. The cancellation in $k_J$ therefore says nothing about $\gamma$ in Eq.~(\ref{eq:deltac_param}). The threshold coefficient remains genuinely undetermined by linear theory.

\subsubsection{Approach 2: Spherical collapse model}

The parameterization of the collapse threshold in this subsection is phenomenological. The coefficient $\gamma$ in Eq.~(\ref{eq:deltac_param}) cannot be determined from linear theory alone; it requires a full numerical simulation of relativistic collapse in Rastall gravity. We treat $\gamma$ as a free $\mathcal{O}(1)$ parameter throughout. Section~5.1.3 provides a brief cross-check via the Misner--Sharp formalism.

For a top-hat overdensity, the modified Friedmann equations give an effective density inside the perturbed region:

\begin{equation}
\rho_{\text{eff}} = \rho\left[1 + \delta\left(1 - \frac{3\lambda}{1-4\lambda}\right)\right]. \label{eq:rhoeff}
\end{equation}

The perturbed scale factor then satisfies:

\begin{equation}
\left(\frac{\dot{a}_p}{a_p}\right)^2 = H^2\left[1 + \delta\left(1 - \frac{3\lambda}{1-4\lambda}\right)\right]. \label{eq:collapse}
\end{equation}

A proper determination of $\delta_c(\lambda)$ would require solving the fully nonlinear relativistic collapse equations in Rastall gravity, which is beyond the scope of this paper. We instead adopt the following phenomenological parameterisation, valid to first order in $\lambda$:

\begin{equation}
\delta_c(\lambda)
=
\delta_c^{\rm GR}
\left( 1 + \gamma \lambda \right),
\label{eq:deltac_param}
\end{equation}

The coefficient $\gamma$ encodes the nonlinear collapse response andfis expected to be $\mathcal{O}(1)$; its precise value requires dedicated numerical simulations in Rastall gravity that have not yet been performed. Positive $\gamma$ raises the threshold for $\lambda > 0$ (harder to collapse), while negative $\lambda$ lowers it. For the abundance calculation in Section~5.3 only the combination $(\gamma - \alpha_\sigma)$ enters, and its sign determines whether PBH production is enhanced or suppressed relative to GR.

\subsubsection{Approach 3: Misner--Sharp cross-check}

As a qualitative cross-check, one may consider the Misner--Sharp formalism \cite{MisnerSharp1964}. A naive substitution $G \to G_{\mathrm{eff}} = G(1 - 4\lambda)$
in the compaction function $C = 2GM_{\mathrm{MS}}/R$ would suggest a shift
$\delta_c(\lambda) \simeq \delta_c^{\mathrm{GR}}(1 + 4\lambda)$ if only the
gravitational sector is modified.

However, this estimate is incomplete. In Rastall gravity, the effective
sound speed is simultaneously modified,
$c_s^2(\lambda) = (1 - 4\lambda)/3$, and as shown in Section 5.1.1 these
two effects cancel at the level of the Jeans criterion. Since the collapse
threshold $\delta_c$ is determined by fully nonlinear dynamics involving both
gravity and pressure gradients, the Misner--Sharp argument cannot be used to
reliably determine the coefficient $\gamma$.

We therefore interpret the Misner--Sharp estimate only as indicating that the
dependence of $\delta_c$ on $\lambda$ is linear to leading order, while leaving
the coefficient $\gamma$ as a genuinely undetermined $\mathcal{O}(1)$ parameter
to be fixed by future nonlinear simulations.

\begin{figure}[htb!]
\centering
\includegraphics[width=0.99\textwidth]{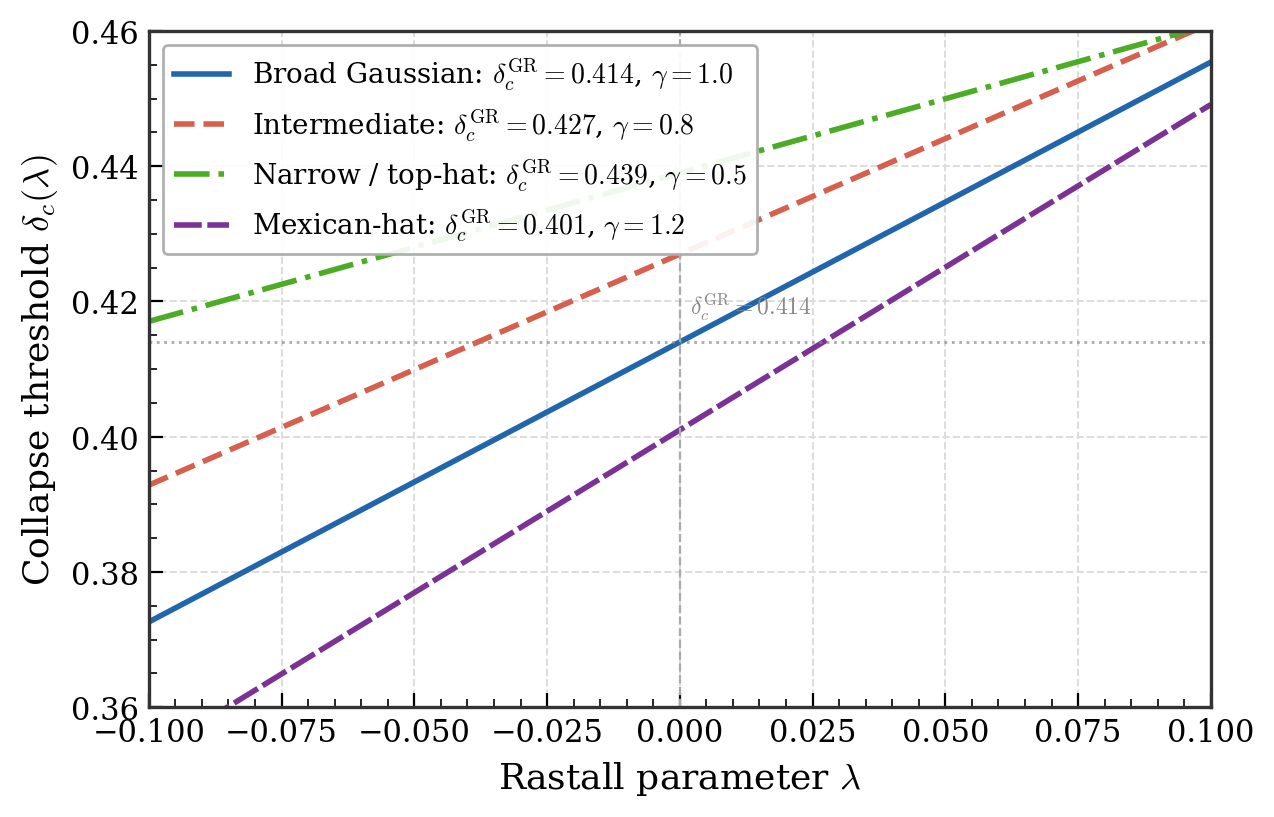}
\caption{Collapse threshold $\delta_c(\lambda)$ for four representative perturbation profile shapes, each characterised by its GR baseline $\delta_c^{\rm GR}$ from numerical simulations \cite{Musco2013,Musco2020,Escriva2020} and a representative $\mathcal{O}(1)$ Rastall correction coefficient $\gamma$. All profiles produce the same qualitative linear shift; the $\gamma$ values are illustrative and not determined from first principles.}
\label{fig2}
\end{figure}

Figure~\ref{fig1} shows $\delta_c(\lambda)=\delta_c^{\rm GR}(1+\gamma\lambda)$ for representative values $\gamma\in\{0,0.5,1,2\}$, together with the $\pm0.025$ GR profile-shape uncertainty band \cite{Musco2020,Escriva2020,Escriva2024}. This band reflects the spread in the GR threshold value $\delta_c^{\rm GR}$ obtained from numerical simulations of relativistic collapse for different perturbation profile shapes: broad quasi-Gaussian profiles give $\delta_c^{\rm GR}\simeq 0.414$, while narrower or more peaked profiles shift this value by up to $\pm 0.025$ \cite{Escriva2020,Escriva2024}. The band therefore represents an irreducible GR uncertainty on the baseline, independent of the Rastall correction. The $\gamma=0$ curve is flat, reflecting the Jeans cancellation derived above. Figure~\ref{fig2} extends this picture to four perturbation profile shapes, each with its own GR baseline $\delta_c^{\rm GR}$ and illustrative $\gamma$ value, showing that the qualitative linear shift persists across all profile types. The key conclusion from both figures is directional: for $\lambda>0$ the threshold can only increase, while for $\lambda<0$ it decreases, independently of the precise $\gamma$.

\subsection{Variance of Density Fluctuations}

The variance of density fluctuations smoothed on scale $R = 1/k$ is:

\begin{equation}
\sigma^2(R) = \int_0^\infty \frac{dk}{k} \mathcal{P}_\delta(k) W^2(kR), \label{eq:sigma}
\end{equation}

with $W(kR)$ a window function. From the perturbation analysis, the density power spectrum at horizon crossing is:

\begin{equation}
\mathcal{P}_\delta(k) = \frac{16}{81}(1 + 2\alpha_{\sigma}\lambda)\mathcal{T}^2(k\tau)\mathcal{P}_\mathcal{R}(k), \label{eq:Pdelta}
\end{equation}

where the coefficient $\alpha_{\sigma}=\alpha_H$ arises from the perturbation matching at horizon crossing (Eq.~(\ref{eq:horcross})), and is independent of the nonlinear collapse coefficient $\gamma$. $\mathcal{T}(k\tau)$ is the transfer function. We use the symbol $\alpha_{\sigma}$ (rather than the conventional $\beta$) to avoid a notation clash with the PBH mass fraction $\beta(M)$ defined in Section~5.3.

For a scale-invariant spectrum $\mathcal{P}_\mathcal{R}(k) = A$, smoothed on the horizon scale $R = 1/(aH)$:

\begin{equation}
\sigma(M_*) = \frac{4}{9}(1 + \alpha_{\sigma}\lambda)\sqrt{A}, \label{eq:sigmaM}
\end{equation}

where $M_*$ is the horizon mass at horizon crossing.

\subsection{Primordial Black Hole Abundance}

We compute the PBH mass fraction using the Press--Schechter formalism. The fraction collapsing into black holes at formation is

We emphasize that the Press--Schechter calculation below employs a Gaussian distribution for the density contrast. This approximation neglects the effects of nonlinearities in the density contrast, which can generate a non-Gaussian tail in the distribution \cite{DeLuca2019, Germani2019b} and modify the abundance estimate by an $\mathcal{O}(1)$ factor in the argument of the exponential. A proper treatment of non-Gaussian statistics would require mapping between the curvature perturbation and the density contrast through the nonlinear relation $\delta = -\frac{2}{3}\left(\frac{k}{aH}\right)^2\Phi + \ldots$ evaluated beyond linear order \cite{DeLuca2019, Germani2019b}. Since we are primarily interested in the relative change in abundance due to the Rastall parameter $\lambda$ rather than the absolute normalization, the Gaussian approximation is sufficient for establishing the exponential sensitivity~(\ref{eq:enhancement}), but this limitation should be borne in mind when interpreting quantitative constraints.

\begin{equation}
\beta(M)
=
\int_{\delta_c}^{\infty}
\frac{d\delta}{\sqrt{2\pi}\,\sigma(M)}
\exp\left(
-\frac{\delta^2}{2\sigma^2(M)}
\right)
=
\frac12
\mathrm{erfc}
\left(
\frac{\delta_c(\lambda)}
{\sqrt{2}\,\sigma(M,\lambda)}
\right),
\label{eq:beta_exact}
\end{equation}

In the rare-peak regime where $\sigma \ll \delta_c$, the erfc is dominated by its exponential tail and Eq.~(\ref{eq:beta_exact}) simplifies to

\begin{equation}
\beta(M)
\approx
\frac{\sigma(M,\lambda)}
{\sqrt{2\pi}\,\delta_c(\lambda)}
\exp\left(
-\frac{\delta_c^2(\lambda)}
{2\sigma^2(M,\lambda)}
\right).
\label{eq:beta_tail}
\end{equation}

Substituting the leading-order Rastall modifications:

\begin{align}
\delta_c(\lambda)
&=
\delta_c^{\rm GR}
\left(1+\gamma\lambda\right), \\
\sigma(M,\lambda)
&=
\sigma_{\rm GR}(M)
\left(1+\alpha_{\sigma}\lambda\right),
\end{align}

with $\gamma$ and $\alpha_\sigma$ as defined above. Define the peak significance

\begin{equation}
\nu_c(\lambda)
\equiv
\frac{\delta_c(\lambda)}
{\sigma(M,\lambda)}.
\end{equation}

\begin{figure}[H]
\centering
\includegraphics[width=0.99\textwidth]{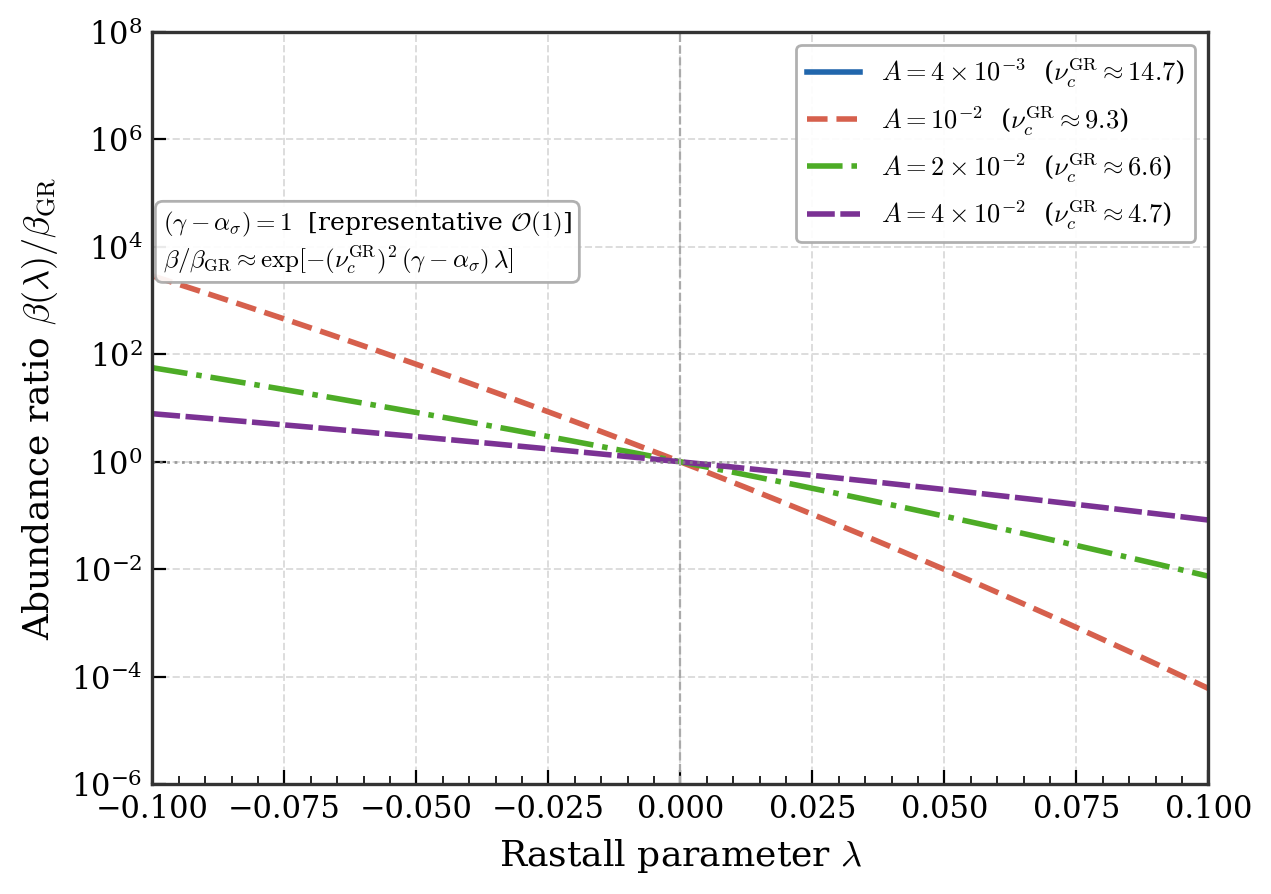}
\caption{PBH abundance ratio $\beta(\lambda)/\beta_{\rm GR}$ as a function of the Rastall parameter~$\lambda$, computed from the exact erfc formula Eq.~(\ref{eq:beta_exact}) for four primordial amplitudes~$A$ (setting the rarity $\nu_c^{\rm GR}=\delta_c^{\rm GR}/\sigma_{\rm GR}$). The curves are shown for a representative $\mathcal{O}(1)$ value $(\gamma-\alpha_{\sigma})=1$; both signs are physically possible, and the opposite sign would reflect the curves about $\lambda=0$. The key feature is the steep exponential dependence on $\nu_c^{\rm GR}$: rare-peak scenarios ($\nu_c^{\rm GR}\gtrsim 9$) suffer orders-of-magnitude changes in abundance for $|\lambda|\lesssim0.05$. The Gaussian approximation caveat of Section~5.3 applies.}
\label{fig3}
\end{figure}

Expanding to first order in $\lambda$:

\begin{equation}
\nu_c(\lambda)
=
\nu_c^{\rm GR}
\left[
1+(\gamma-\alpha_{\sigma})\lambda
\right],
\label{eq:nu_expansion}
\end{equation}

where $\nu_c^{\rm GR} = \delta_c^{\rm GR}/\sigma_{\rm GR}$. Substituting into Eq.~(\ref{eq:beta_tail}):

\begin{equation}
\frac{\beta(\lambda)}{\beta_{\rm GR}}
\approx
\exp\left[
- (\nu_c^{\rm GR})^2
(\gamma-\alpha_{\sigma})\lambda
\right],
\label{eq:enhancement}
\end{equation}

where $\beta_{\rm GR}$ is the GR abundance. The factor $(\nu_c^{\rm GR})^2$ amplifies the effect: for the rare-peak scenarios relevant to PBH dark matter, $\nu_c^{\rm GR} \sim 5$--$10$, so even $|\lambda| \sim 0.05$ shifts the abundance by several orders of magnitude. This is illustrated in Figure~\ref{fig3}, which shows $\beta(\lambda)/\beta_{\rm GR}$ for four representative primordial amplitudes $A$ and a representative $\mathcal{O}(1)$ value $(\gamma-\alpha_{\sigma})=1$. Both signs of $(\gamma-\alpha_{\sigma})$ are physically possible; the figure shows one representative case. The steeper the rare-peak hierarchy ($\nu_c^{\rm GR}$ larger), the more dramatic the effect.

It is instructive to note that even the conservative limit $\gamma=0$ (unchanged collapse threshold, consistent with the Jeans cancellation of Section~5.1.1) gives a non-trivial abundance modification through $\alpha_\sigma$ alone:
\begin{equation}
\left.\frac{\beta(\lambda)}{\beta_{\rm GR}}\right|_{\gamma=0}
\approx
\exp\!\left[(\nu_c^{\rm GR})^2\,\alpha_{\sigma}\lambda\right].
\label{eq:enhancement_gamma0}
\end{equation}
Since $\alpha_\sigma$ is $\mathcal{O}(1)$ and Figure~\ref{fig4} confirms it is negative (positive $\lambda$ reduces the horizon-crossing amplitude), this expression gives exponential sensitivity to $\lambda$ through the variance modification alone, independently of $\gamma$.

\subsection{Numerical Solution of the Master Equation: Combined Rastall Effects}
\label{sec:fig4}

\begin{figure}[htb!]
\centering
\includegraphics[width=0.99\textwidth]{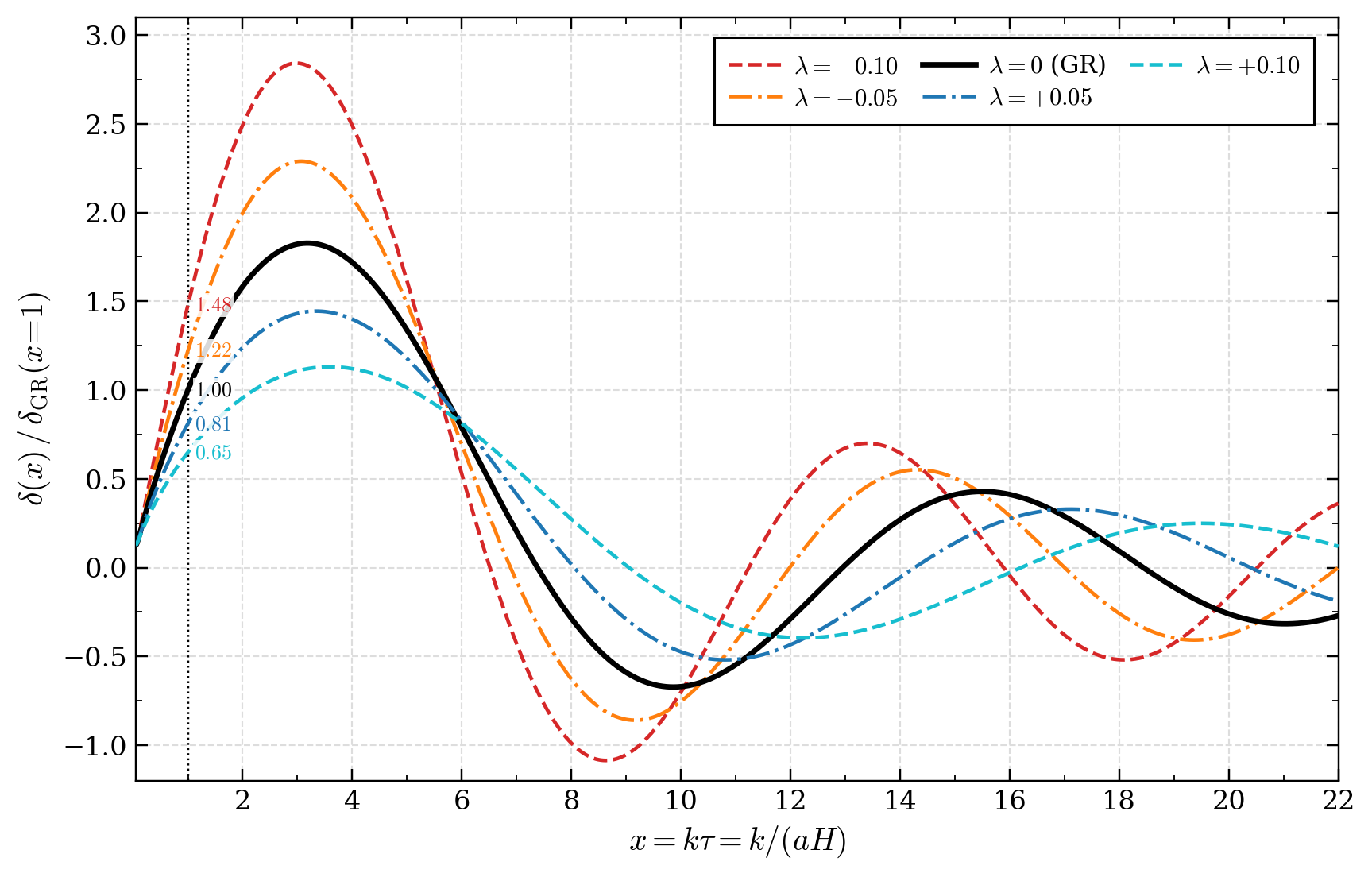}
\caption{Numerical solution of the Rastall master
equation~(\ref{eq:masterx}), showing the normalised density contrast
$\delta(x)/\delta_{\rm GR}(x=1)$ for five values of the Rastall
parameter~$\lambda$. The dotted vertical line marks horizon crossing
($x = 1$). In the super-horizon regime ($x < 1$) all curves are
nearly indistinguishable, consistent with the $\mathcal{O}(x^2)$
correction in Eq.~(\ref{eq:superhor}). At horizon crossing the
annotated ratios $\delta_H/\delta_H^{\rm GR} \in [0.65,\,1.48]$
directly encode the $(1+\alpha_H\lambda)$ modification of
Eq.~(\ref{eq:horcross}): negative $\lambda$ enhances the
horizon-crossing amplitude while positive $\lambda$ suppresses it,
confirming that $\alpha_\sigma = \alpha_H < 0$. In the sub-horizon
regime ($x > 1$) the curves oscillate with the modified sound speed
$c_s(\lambda) = \sqrt{(1-4\lambda)/3}$, producing phase-shifted
oscillations whose first-zero locations shift earlier for
$\lambda > 0$ and later for $\lambda < 0$.}
\label{fig4}
\end{figure}

Figure~\ref{fig4} shows the numerical solution of the master equation~(\ref{eq:masterx}) normalised by the GR horizon-crossing value $\delta_{\rm GR}(x=1)$, and brings together the three Rastall effects discussed in Sections~3--5.

Three physically distinct effects are simultaneously visible. First, in the \emph{super-horizon} regime ($x<1$), all curves track each other closely, consistent with the tiny $\mathcal{O}(x^2)$ correction in Eq.~(\ref{eq:superhor}). Second, at \emph{horizon crossing} ($x=1$), the curves reach different amplitudes $\delta_H/\delta_H^{\rm GR}\in[0.65,1.48]$ for $\lambda\in[-0.10,+0.10]$, directly illustrating the $(1+\alpha_H\lambda)$ modification of Eq.~(\ref{eq:horcross}) that feeds into the variance $\sigma(M,\lambda)$. The horizon-crossing amplitude is reduced for positive $\lambda$ (reduced sound speed $c_s^2=(1-4\lambda)/3$) and enhanced for negative $\lambda$, confirming that $\alpha_\sigma=\alpha_H$ is negative. Third, in the \emph{sub-horizon} regime ($x>1$), the curves oscillate with sound speed $c_s(\lambda)=\sqrt{(1-4\lambda)/3}$, producing a measurable phase shift and shifted first-zero locations (earlier for $\lambda<0$, later for $\lambda>0$).

The precise numerical value of $\alpha_\sigma$ requires a dedicated semi-analytic matching calculation beyond the scope of this work. The sign (negative) and order of magnitude ($|\alpha_\sigma|=\mathcal{O}(1)$) are established by the numerical solution; a precise value would require solving the matching problem at sub-horizon scales analytically.

\section{Implications for PBH Dark Matter Scenarios}

Microlensing surveys, CMB spectral distortions, and gravitational-wave observations jointly constrain the PBH abundance across most of the mass range, leaving a handful of windows where PBHs could make up a significant---possibly dominant---fraction of dark matter: roughly the asteroid-mass window ($10^{17}$--$10^{22}$~g), the lunar-mass window ($10^{20}$--$10^{24}$~g), and parts of the stellar range. Comprehensive reviews of these constraints are given in Refs.~\cite{Carr:2020gox, Green2021, Niikura2019, Tisserand2007, Carr:2009jm}.

In GR this requires the primordial curvature power spectrum to reach $\mathcal{P}_{\mathcal R} \sim 10^{-2}$--$10^{-1}$ at small scales, several orders of magnitude above its CMB value $\simeq 2\times10^{-9}$. The degree of fine-tuning involved is one of the recurring concerns about the PBH dark matter scenario.

In Rastall gravity the background during radiation domination is, as we have shown, exactly that of GR. But the perturbation dynamics carry $\lambda$-dependent corrections, and through the exponential dependence of $\beta$ on $\nu_c^2$ these translate into large changes in the abundance. Specifically:
\begin{equation}
\nu_c(\lambda)
=
\nu_c^{\rm GR}
\left[1+(\gamma-\alpha_{\sigma})\lambda\right],
\end{equation}
where $\alpha_\sigma$ and $\gamma$ are the variance and threshold coefficients derived above. Since $\beta \sim \exp(-\nu_c^2/2)$, the exponential amplification means that:

\begin{itemize}
\item If $(\gamma-\alpha_{\sigma})\lambda < 0$, the effective threshold $\nu_c$ is reduced and PBH formation is exponentially enhanced.
\item If $(\gamma-\alpha_{\sigma})\lambda > 0$, $\nu_c$ increases and PBH formation is suppressed.
\end{itemize}

Whether formation is enhanced or suppressed therefore depends on the relative sign of $\gamma$ and $\alpha_\sigma$---Rastall gravity does not produce a universal enhancement.

For rare-peak scenarios with $\nu_c^{\rm GR} \sim 5$--$10$, a percent-level shift in $(\gamma-\alpha_\sigma)\lambda$ translates into order-of-magnitude changes in $\beta$, while shifting the required primordial amplitude by only $\mathcal{O}(10\%)$. The fine-tuning demands on the inflationary power spectrum can therefore be substantially relaxed---or, if the effect goes the other way, tightened---depending on the sign of $(\gamma-\alpha_\sigma)$.

Turning the argument around, any observational bound on the PBH abundance immediately becomes a bound on the combination $(\gamma-\alpha_\sigma)\lambda$. Since the expansion history is unchanged, such a bound isolates perturbation-level physics cleanly.

We should be clear about what is and is not specific to Rastall gravity here. The exponential sensitivity of $\beta$ to $\nu_c$ is a generic feature of rare-event statistics and would arise in any theory that shifts $\nu_c$ by an $\mathcal{O}(1)$ amount. What is specific to Rastall gravity is threefold. First, the variance correction $\alpha_\sigma$ is determined entirely by the linear master equation---there is no free parameter in the variance; its sign is fixed negative by the perturbation dynamics, as confirmed numerically in Figure~\ref{fig4}. Second, the Jeans cancellation (Section~5.1.1) is a non-trivial structural consequence of the particular coupling $\nabla_\mu T^{\mu\nu} = \lambda \nabla^\nu R$: the same factor $(1-4\lambda)$ appears in both $G_{\rm eff}$ and $c_s^2$, so the Jeans scale is unaffected even though gravity and pressure are each modified. This cancellation does not hold generically in other modified gravity theories. Third, and most importantly, the background expansion is \emph{exactly} GR for any $\lambda$---not approximately, not at leading order, but exactly---so there is no degeneracy between Rastall corrections and a modified Hubble history. PBH bounds therefore probe $\lambda$ in a regime where every other cosmological background observable is silent.

If a PBH dark matter component with a measured mass function were to be established, it would also become possible to infer the relationship between the primordial amplitude and the observed abundance, which in Rastall gravity carries a distinctive $\lambda$-dependent form. PBHs therefore complement CMB and large-scale structure as probes of small departures from GR.

\section{Conclusions and Outlook}

We have studied PBH formation in Rastall gravity during radiation domination. The main finding is that, although the background expansion of a pure radiation fluid is identical to GR at all orders, the perturbation dynamics differ starting at linear order in $\lambda$.

The master equation for the density contrast shows that both the effective gravitational coupling and the sound speed are modified at $\mathcal{O}(\lambda)$. These two corrections happen to cancel in the Jeans wavenumber, leaving the linear stability criterion unchanged from GR. The horizon-crossing amplitude, however, does shift, and as a consequence the variance of density perturbations acquires a Rastall correction:
\begin{equation}
\sigma(M,\lambda)
=
\sigma_{\rm GR}(M)
\left(1+\alpha_{\sigma}\lambda\right),
\end{equation}
where $\alpha_\sigma$ is fixed by the linear perturbation theory (Eqs.~(\ref{eq:Pdelta})--(\ref{eq:sigmaM})). This is the primary first-principles prediction of the paper: it follows directly from linear perturbation theory and is entirely independent of the unknown collapse coefficient $\gamma$.

The nonlinear collapse threshold cannot be determined from linear theory. In the absence of relativistic collapse simulations in Rastall gravity, we used the phenomenological ansatz (Eq.~(\ref{eq:deltac_param}))
\begin{equation}
\delta_c(\lambda)
=
\delta_c^{\rm GR}
\left(1+\gamma\lambda\right),
\end{equation}
with $\gamma$ an unknown $\mathcal{O}(1)$ coefficient. Together, these give (Eq.~(\ref{eq:nu_expansion})):
\begin{equation}
\nu_c(\lambda)
=
\nu_c^{\rm GR}
\left[1+(\gamma-\alpha_{\sigma})\lambda\right].
\end{equation}
Because $\beta \propto \exp(-\nu_c^2/2)$, the exponential sensitivity turns even modest $\lambda$ into large changes in the PBH abundance (Eq.~(\ref{eq:enhancement})):
\begin{equation}
\frac{\beta(\lambda)}{\beta_{\rm GR}}
\sim
\exp\!\left[
-(\nu_c^{\rm GR})^2(\gamma-\alpha_{\sigma})\lambda
\right].
\end{equation}

The two contributions to the abundance modification have distinct theoretical status. The variance coefficient $\alpha_{\sigma}$ is determined by the linear perturbation analysis of Sections~3 and~4 and is a first-principles prediction: Figure~\ref{fig4} confirms it is negative (positive $\lambda$ reduces the horizon-crossing amplitude) and $\mathcal{O}(1)$ in magnitude, though its precise value requires a dedicated matching calculation. The threshold coefficient $\gamma$ is a phenomenological parameter encoding nonlinear collapse dynamics that have not yet been simulated. The sign of $(\gamma-\alpha_{\sigma})$ determines the direction of the effect; both enhancement and suppression are physically possible. Even in the conservative limit $\gamma=0$---consistent with the Jeans cancellation---the abundance remains exponentially sensitive to $\lambda$ through $\alpha_\sigma$ alone (Eq.~(\ref{eq:enhancement_gamma0})), so the central conclusion is not contingent on the unknown value of $\gamma$.

The upshot is that PBHs probe perturbation-level modifications of gravity even when the background expansion is completely standard. The exponential sensitivity of the abundance to $\nu_c$ is what makes PBH constraints so powerful for theories of this kind.

\subsection*{Outlook}

The most pressing extension is a numerical simulation of relativistic collapse in Rastall gravity using, for instance, a Hernandez--Misner or Misner--Sharp code adapted to the modified field equations. This would fix $\gamma$ from first principles and remove the main remaining theoretical uncertainty \cite{Musco2013, Musco2020, Escriva2020, Escriva2024}.

On the perturbation theory side, a full gauge-invariant treatment of curvature perturbations in Rastall gravity would be worthwhile. Non-minimal coupling between matter and geometry generically generates non-adiabatic modes, and their role in the horizon-crossing amplitude has not been assessed ---such a treatment would also allow direct comparison in the compaction-function language of Ref.~\cite{Escriva2020}.

The Gaussian approximation used in the abundance calculation is another obvious target for improvement. The nonlinear relation between the curvature perturbation and the density contrast generates a non-Gaussian tail \cite{DeLuca2019, Germani2019b} that can shift the abundance by an $\mathcal{O}(1)$ factor in the exponent; incorporating this using the methods of Refs.~\cite{DeLuca2019, Germani2019b} would give a more reliable quantitative estimate.

It would also be interesting to calculate the induced stochastic gravitational wave background sourced by the enhanced scalar perturbations \cite{Bagui2025}. Since the background expansion is unmodified, any signal would be a clean signature of perturbation-level physics, and the prospects for detection with LISA and PTA experiments deserve attention.

Finally, a self-consistent calculation of the primordial power spectrum within a specific Rastall inflationary model would allow one to predict $\lambda$-dependent spectral features directly, rather than treating $\mathcal{P}_\mathcal{R}(k)$ as input \cite{FineTuning2024a, FineTuning2024b}.

In summary, PBH formation in Rastall gravity is sensitive to perturbation-level departures from GR even when the background expansion is identical. The exponential sensitivity of the abundance to $\lambda$ makes future PBH observations a promising complement to CMB and large-scale structure tests of the theory \cite{Carr:2020gox, Green2021, Niikura2019, Tisserand2007}. We caution that the present results assume a broad perturbation profile and a Gaussian distribution for the density contrast; relaxing either assumption is a natural next step.

\section*{Acknowledgments}

The author would like to thank Prof. M. Sami and Dr. Debottam Nandi for insightful comments during the preparation of the manuscript and Sourav Mridha for carefully going through the manuscript and helping to typeset. The author would like to thank the anonymous referees for thier invaluable comments. The work of MRG is supported by the Science and Engineering Research Board (SERB), DST, Government of India, under the Grant Agreement number CRG/2022/004120 (Core Research Grant). MRG would like to thank IUCAA, Pune, India, for their hospitality during the visit as an associate, where this work was initiated.

\end{document}